\documentclass{article}

\usepackage{amssymb}
\usepackage{amsfonts}
\usepackage{epsfig}

\newcommand{\Au}{{\bf (A.1)}}
\newcommand{\Ad}{{\bf (A.2)}}

\newcommand{\qchap}{\hat{x}_{p_n}}
\newcommand{\qvrai}{x_{p_n}}
\newcommand{\thetachap}{\hat{\theta}_n}

\newcommand{\unchap}{X_{n-k_n+1,n}}
\newcommand{\somi}{\sum_{i=1}^{k-1}}

\newcommand{\invH}{H^{\leftarrow}}
\newcommand{\invF}{F^{-1}}
\newcommand{\toP}{\stackrel{P}{\to}}
\newcommand{\tod}{\stackrel{d}{\to}}
\newcommand{\simP}{\stackrel{P}{\sim}}

\newcommand{\egl}{\stackrel{d}{=}}
\newcommand{\R}{\mathbb{R}}
\newcommand{\E}{\mathbb{E}}
\newcommand{\ld}[1] {\log_2\left({#1}\right)}

\newcommand{\CQFD}
{%
\mbox{}%
\nolinebreak%
\hfill%
\rule{2mm}{2mm}%
\medbreak%
\par%
}

\newtheorem{Theo}{Theorem}
\newtheorem{prop}{Proposition}
\newtheorem{Coro}{Corollary}
\newtheorem{Lem}{Lemma}

\begin{document}

\title{Comparison of  Weibull tail-coefficient estimators}
\author{Laurent Gardes$^{(1)}$ \& St\'ephane Girard$^{(2,\star)}$}
\date{$^{(1)}$Universit\'e Grenoble 2, LabSAD, BP 47, 38040 Grenoble cedex 9, France, {\tt Laurent.Gardes@upmf-grenoble.fr}\\
$^{(2,\star)}$Universit\'e Grenoble 1, LMC-IMAG, BP 53, 38041 Grenoble cedex 9, France, {\tt Stephane.Girard@imag.fr}
(corresponding author)}

\maketitle
\begin{abstract}
We address the problem of estimating the Weibull tail-coefficient
which is the regular variation exponent of the inverse failure rate
function.  We propose a family of estimators of this coefficient
and an associate extreme quantile estimator.  Their
asymptotic normality are established and their asymptotic
mean-square errors are compared.  The results are illustrated
on some finite sample situations.  \\


\end{abstract}

\section{Introduction}
\label{intro}

Let $X_1, X_2,\dots,X_n$ be a sequence of independent and identically
distributed random variables with cumulative distribution function $F$.
We denote by $X_{1,n} \leq \ldots \leq X_{n,n}$ their associated order statistics.
We address the problem of estimating the Weibull tail-coefficient $\theta>0$
defined when the distribution tail satisfies

\begin{description}
\item [(A.1)] $1-F(x)=\exp(-H(x))$, $x\geq x_0\geq 0$, 
$\invH(t)=\inf\{x,\; H(x)\geq t\}=t^\theta\ell(t)$,
\end{description}
\noindent where $\ell$ is a slowly varying function {\it i.e.} 
$$
\ell(\lambda x)/\ell(x)\to 1 \mbox{ as } x\to\infty \mbox{ for all } \lambda>0.
$$
The inverse cumulative hazard function $\invH$ is said to be 
regularly varying at infinity with index~$\theta$
and this property is denoted by $\invH\in{\mathcal R}_\theta$,
see~\cite{bingolteu87} for more details on this topic.  
As a comparison, Pareto type distributions satisfy
$(1/(1-F))^\leftarrow \in{\mathcal R}_\gamma$, and
$\gamma>0$ is the so-called extreme value index.
Weibull tail-distributions include for instance
Gamma, Gaussian and, of course, Weibull distributions.  

\noindent Let $(k_n)$ be a sequence of integers such that $1 \leq k_n < n$ and 
$(T_n)$ be a positive sequence. We examine the asymptotic behavior of
the following family of estimators of $\theta:$
\begin{equation}
\label{deftheta}
\hat{\theta}_n = \frac{1}{T_n} \frac{1}{k_n} \sum_{i=1}^{k_n} (\log(X_{n-i+1,n})-\log(X_{n-k_n+1,n})).
\end{equation}
Following the ideas of~\cite{quantWT}, an estimator of the
extreme quantile $\qvrai$ can be deduced from~(\ref{deftheta}) by:
\begin{equation}
\label{defqchap}
\qchap = \unchap \left(\frac{\log (1/p_n)}{\log (n/k_n)}\right)^{\thetachap}
=: \unchap \tau_n^{\thetachap}.  
\end{equation}
Recall that an extreme quantile $\qvrai$ of order $p_n$ is defined
by the equation
$$
1-F(\qvrai)=p_n, \mbox{ with } 0<p_n<1/n.
$$
The condition $p_n<1/n$ is very important in this context. It usually implies
that $\qvrai$ is larger than the maximum observation of the sample.  
This necessity to extrapolate sample results to areas where no data
are observed occurs in reliability~\cite{Ditle}, hydrology~\cite{Smith},
finance~\cite{EMBR},...
We establish in Section~\ref{AN} the asymptotic normality of $\thetachap$
and $\qchap$.  The asymptotic mean-square error of some particular
members of~(\ref{deftheta}) are compared in Section~\ref{examples}.
In particular, it is shown that family~(\ref{deftheta}) encompasses the 
estimator introduced in~\cite{thetaWT} and
denoted by $\hat\theta_n^{(2)}$ in the sequel.
In this paper,
the asymptotic normality of $\hat\theta_n^{(2)}$ 
is obtained under weaker conditions.  
Furthermore, we show that other members of family~(\ref{deftheta}) should be
preferred in some typical situations. 
We also quote some other estimators of $\theta$ which do not
belong to family~(\ref{deftheta}): \cite{BeirBro,BeirBouq,BRO,KLUP}.
We refer to~\cite{thetaWT} for a comparison with $\hat\theta_n^{(2)}$.  
The asymptotic results are illustrated in Section~\ref{simul} on finite
sample situations. Proofs are postponed to Section~\ref{proofs}.

\section{Asymptotic normality}
\label{AN}

To establish the asymptotic normality of $\hat{\theta}_n$, we need a second-order condition on $\ell$:
\begin{description}
\item [(A.2)] 
There exist $\rho\leq 0$ and $b(x)\to 0$ such that
uniformly locally on $\lambda\geq 1$ 
$$
 \log\left(\frac{\ell(\lambda x)}{\ell(x)}\right) \sim b(x) K_\rho(\lambda), \mbox{ when }
x\to\infty, 
$$
with $K_\rho(\lambda)=\int_1^\lambda u^{\rho-1} du$.
\end{description}
It can be shown~\cite{GEL} that necessarily $|b|\in{\mathcal R}_\rho$.
The second order parameter $\rho\leq 0$ tunes the rate of convergence
of $\ell(\lambda x)/\ell(x)$ to 1. The closer $\rho$ is to 0, the slower
is the convergence. Condition \Ad~is the cornerstone in all proofs of
asymptotic normality for extreme value estimators.
It is used in~\cite{Hill,Hausler,BEIR} to prove the asymptotic normality of
estimators of the extreme value index~$\gamma$. 
In regular case, as noted in~\cite{Gomes3}, one can choose
$b(x)={x\ell'(x)}/{\ell(x)}$ leading to
\begin{equation}
\label{defb}
b(x)= \frac{x e^{-x}}{\invF(1-e^{-x})f(\invF(1-e^{-x}))}-\theta,
\end{equation}
where $f$ is the density function associated to $F$. \\ 
\noindent Let us introduce the following functions : for $t >0$ and $\rho \leq 0$,
\begin{eqnarray*}
 \mu_{\rho}(t) &=&  \int_0^{\infty} K_{\rho} \left ( 1+\frac{x}{t} \right ) {\rm{e}}^{-x}dx\\
 \sigma_{\rho}^2(t)&=& \int_0^{\infty}  K_{\rho}^2 \left ( 1+\frac{x}{t} \right ) {\rm{e}}^{-x}dx- \mu^2_{\rho}(t), 
\end{eqnarray*}
and let $a_n=\mu_{0}(\log(n/k_n))/T_n-1$.
\noindent As a preliminary result, we propose an asymptotic expansion of $(\hat{\theta}_n - \theta )$:
\begin{prop}
\label{decomptheta}
Suppose \Au~and \Ad~hold. If $k_n \to \infty$, $k_n/n \to 0$, $T_n \log(n/k_n)\to 1$ and $k_n^{1/2}b(\log(n/k_n)) \to \lambda \in \R$ then,
\begin{eqnarray*}
k_n^{1/2} (\hat{\theta}_n - \theta ) & = & \theta \xi_{n,1} + 
 {\theta}{\mu_0(\log(n/k_n))} \xi_{n,2} + k_n^{1/2} \theta a_n \\
 \ & + & k_n^{1/2}b(\log(n/k_n))(1+o_{\rm{P}}(1)),
\end{eqnarray*}
where $\xi_{n,1}$ and  $\xi_{n,2}$ converge in distribution to a standard normal distribution.
\end{prop}
Similar distributional representations exist for various estimators
of the extreme value index $\gamma$.  They are used in~\cite{HP}
to compare the asymptotic properties of several tail index estimators.  
In~\cite{Gomes1}, a bootstrap selection of $k_n$ is derived from 
such a representation.  It is also possible to derive bias reduction
method as in~\cite{Gomes2}.
The asymptotic normality of $\hat{\theta}_n$ is a straightforward consequence of Proposition~\ref{decomptheta}.
\begin{Theo}
\label{NAtheta}
Suppose \Au~and \Ad~hold. If $k_n \to \infty$, $k_n/n \to 0$, $T_n \log(n/k_n)\to 1$ and $k_n^{1/2}b(\log(n/k_n)) \to \lambda \in \R$ then,
\[  k_n^{1/2} ({\hat{\theta}}_n-\theta-b(\log(n/k_n)) - \theta a_n ) \tod {\mathcal{N}}(0,\theta^2). \]
\end{Theo}

\noindent Theorem \ref{NAtheta} implies that the Asymptotic Mean Square Error (AMSE) of ${\hat{\theta}}_n$ is given by :
\begin{equation}
\label{AMSE}
AMSE({\hat{\theta}}_n) = (\theta a_n + b(\log(n/k_n)) )^2 + \frac{\theta^2}{k_n}.
\end{equation}
\noindent It appears that all estimators of family~(\ref{deftheta})
share the same variance.  The bias depends on two terms
$b(\log (n/k_n))$ and $\theta a_n$.  A good choice of $T_n$ (depending on the function $b$) could lead to a sequence $a_n$ cancelling the bias.  
Of course, in the general case, the function $b$ is unknown
making difficult the choice of a ``universal'' sequence $T_n$.  
This is discussed in the next section.  

\noindent Clearly, the best rate of convergence in Theorem \ref{NAtheta}
is obtained by choosing $\lambda \neq 0$. In this case, the expression of the intermediate sequence $(k_n)$ is known.

\begin{prop}
\label{vitesse}
If $k_n \to \infty$, $k_n/n \to 0$ and $k_n^{1/2}b(\log(n/k_n)) \to \lambda \neq 0$,
\[ k_n \sim \left ( \frac{\lambda}{b(\log(n))} \right )^2 = \lambda^2(\log(n))^{-2\rho} L(\log(n)), \]
where $L$ is a slowly varying function.
\end{prop}
The ``optimal'' rate of convergence is thus of order $(\log(n))^{-\rho}$,
which is entirely determined by the second order parameter $\rho$:
small values of $|\rho|$ yield slow convergence.  
The asymptotic normality of the extreme quantile estimator~(\ref{defqchap})
can be deduced from Theorem~\ref{NAtheta}:
\begin{Theo}
\label{NAquant}
Suppose \Au~and \Ad~hold. If moreover, $k_n \to \infty$, $k_n/n \to 0$, $T_n \log(n/k_n)\to 1$, $k_n^{1/2}b(\log(n/k_n)) \to 0$ and
\begin{equation}
\label{condcons}
1\leq\lim\inf\tau_n\leq\lim\sup\tau_n<\infty
\end{equation}
then,
\[  \frac{k_n^{1/2}}{\log \tau_n} \left(\frac{\qchap}{\qvrai}-\tau_n^{\theta a_n} \right) \tod {\mathcal{N}}(0,\theta^2). \]
\end{Theo}

\section{Comparison of some estimators}
\label{examples}

First, we propose some choices of the sequence $(T_n)$ leading to different estimators of the Weibull tail-coefficient. 
Their asymptotic distributions are provided, and their AMSE are compared.

\subsection{Some examples of estimators}

-- The natural choice is clearly to take
\[ T_n = T_n^{(1)} =: \mu_0(\log(n/k_n)), \]
in order to cancel the bias term $a_n$.
This choice leads to a new estimator of $\theta$ defined by :
\[
\hat{\theta}_n^{(1)} = \frac{1}{\mu_0(\log(n/k_n))} \frac{1}{k_n} \sum_{i=1}^{k_n} (\log(X_{n-i+1,n})-\log(X_{n-k_n+1,n})).
\]
Remarking that
$$
\mu_{\rho}(t) = {\rm{e}}^{t}  \int_1^{\infty} {\rm{e}}^{-tu}{u^{\rho-1}}du
$$
provides a simple computation method for $\mu_0(\log(n/k_n))$  using 
the Exponential Integral (EI), see for instance~\cite{abrste72}, Chapter~5, pages~225--233.

\noindent -- Girard~\cite{thetaWT} proposes the following estimator of the Weibull tail-coefficient: \[
\hat{\theta}_n^{(2)} =
{\displaystyle\sum_{i=1}^{k_n} (\log(X_{n-i+1,n})  - \log(\unchap)) } \left/
{\displaystyle\sum_{i=1}^{k_n} (\ld{n/i} - \ld{n/k_n}) } \right.,
\]
where $\ld{x} =\log(\log(x))$, $x>1$. Here, we have
\[ T_n = T_n^{(2)} =: \frac{1}{k_n} \sum_{i=1}^{k_n} \log \left ( 1-\frac{\log(i/k_n)}{\log(n/k_n)}\right ). \]
It is interesting to remark that $T_n^{(2)}$ is a Riemann's sum approximation 
of $\mu_{0}(\log(n/k_n))$ since an integration by parts yields:
\[ \mu_0(t)=\int_0^1 \log \left( 1-\frac{\log(x)}{t} \right ) dx. \]

\noindent -- Finally, choosing $T_n$ as the asymptotic equivalent of $\mu_0(\log(n/k_n))$,
$$
T_n=T_n^{(3)}=:1/\log(n/k_n)
$$
 leads to the estimator :
\[
\hat{\theta}_n^{(3)} = \frac{\log(n/k_n)}{k_n} \sum_{i=1}^{k_n} (\log(X_{n-i+1,n})-\log(X_{n-k_n+1,n})).
\]
For $i=1,2,3$, let us denote by $\qchap^{(i)}$ the extreme quantile estimator 
built on $\thetachap^{(i)}$ by ~(\ref{defqchap}). 
Asymptotic normality of these estimators is derived from 
Theorem~\ref{NAtheta} and Theorem~\ref{NAquant}.
To this end, we introduce the following conditions:
\begin{description}
\item [(C.1)] $k_n/n\to 0$,
\item [(C.2)] $\log(k_n)/\log(n)\to 0$,
\item [(C.3)] $k_n/n\to 0$ and $k_n^{1/2}/\log(n/k_n) \to 0$.
\end{description}
Our result is the following:
\begin{Coro}
\label{CorNAtroistheta}
Suppose \Au~and~\Ad~hold, $k_n \to \infty$ and $k_n^{1/2}b(\log(n/k_n)) \to 0$. 
For $i=1,2,3$:
\begin{itemize}
\item[i)] If {\bf (C.i)} hold then
$$
k_n^{1/2}({\hat{\theta}}_n^{(i)} - \theta) \tod {\cal{N}}(0,\theta^2).
$$
\item[ii)] If {\bf (C.i)} and~(\ref{condcons}) hold, then
\[  \frac{k_n^{1/2}}{\log \tau_n} \left(\frac{\qchap^{(i)}}{\qvrai}-1 \right) \tod {\mathcal{N}}(0,\theta^2). \]
\end{itemize}
\end{Coro}
In view of this corollary, the asymptotic normality of ${\hat{\theta}}_n^{(1)}$ is obtained under weaker conditions
than ${\hat{\theta}}_n^{(2)}$ and  ${\hat{\theta}}_n^{(3)}$, since
{\bf (C.2)} implies {\bf (C.1)}. Let us also highlight that  
the asymptotic distribution of ${\hat{\theta}}_n^{(2)}$ is obtained under less
assumptions than in \cite{thetaWT}, Theorem~2, the condition $k_n^{1/2}/\log(n/k_n) \to 0$ being not necessary here.
Finally, note that, if $b$ is not ultimately zero, condition $k_n^{1/2}b(\log(n/k_n)) \to 0$
implies {\bf (C.2)}  (see Lemma~\ref{new}).

\subsection{Comparison of the AMSE of the estimators}

We use the expression of the AMSE given in (\ref{AMSE}) to compare the estimators proposed previously.

\newpage

\begin{Theo}
\label{theoAMSE}
Suppose \Au~and \Ad~hold, $k_n \to \infty$, $\log(k_n)/\log(n) \to 0$ and $k_n^{1/2}b(\log(n/k_n)) \to \lambda \in \R$. 
Several situations are possible:

\begin{itemize}
\item[i)] $b$ is ultimately non-positive. Let us introduce 
$\alpha= -4 \displaystyle\lim_{n\to\infty} b(\log n)\frac{k_n}{\log k_n} \in[0,+\infty]$.
 \ \\
\indent If $\alpha>\theta$, then, for $n$ large enough, $$AMSE({\hat{\theta}}_n^{(2)})<AMSE({\hat{\theta}}_n^{(1)})<AMSE({\hat{\theta}}_n^{(3)}).$$
\indent If $\alpha<\theta$, then, for $n$ large enough, $$AMSE({\hat{\theta}}_n^{(1)}) < \min(AMSE({\hat{\theta}}_n^{(2)}),AMSE({\hat{\theta}}_n^{(3)})).$$
\item[ii)] $b$ is ultimately non-negative. Let us introduce $\beta= 2\displaystyle\lim_{x\to\infty} xb(x) \in[0,+\infty]$.\\
 \ \\
\indent If $\beta > \theta$ then, for $n$ large enough, $$AMSE({\hat{\theta}}_n^{(3)})<AMSE({\hat{\theta}}_n^{(1)})<AMSE({\hat{\theta}}_n^{(2)}).$$
\indent If $\beta < \theta$ then, for $n$ large enough, $$AMSE({\hat{\theta}}_n^{(1)}) < \min(AMSE({\hat{\theta}}_n^{(2)}),AMSE({\hat{\theta}}_n^{(3)})).$$
\end{itemize}
\end{Theo}
It appears that, when $b$ is ultimately non-negative (case ii)), the conclusion does not depend
on the sequence $(k_n)$.   The relative performances of the estimators is 
entirely determined by the nature of the distribution: $\hat\theta_n^{(1)}$ 
has the best behavior, in terms of AMSE, for distributions close to the Weibull
distribution (small $b$ and thus, small $\beta$). At the opposite, $\hat\theta_n^{(3)}$ 
should be preferred for distributions far from the Weibull distribution.  

\noindent The case when $b$ is ultimately non-positive (case i)) is different.  
The value of $\alpha$ depends on $k_n$, and thus, for any distribution, one can obtain
$\alpha=0$ by choosing small values of $k_n $(for instance $k_n=-1/b(\log n)$)
as well as $\alpha=+\infty$ by choosing large values of $k_n$
(for instance $k_n=(1/b(\log n))^{2}$ as in Proposition~\ref{vitesse}).

\section{Numerical experiments}
\label{simul}

\subsection{Examples of Weibull tail-distributions}
Let us give some examples of distributions satisfying
assumptions \Au~and \Ad.

\paragraph{\bf Absolute Gaussian distribution}$|{\mathcal N}(\mu,\sigma^2)|$,
 $\sigma>0$.
From~\cite{EMBR}, Table 3.4.4, we have $\invH(x)=x^{\theta}\ell(x)$,
where  $\theta=1/2$ 
and an asymptotic expansion of the slowly varying function is given by:
$$
\ell(x)=  2^{1/2} \sigma - \frac{\sigma}{2^{3/2}} \frac{\log x}{x} + O(1/x).
$$
Therefore $\rho=-1$ and $b(x)=\log(x)/(4x) + O(1/x)$.
$b$ is ultimately positive, which corresponds to case ii) of
Theorem~\ref{theoAMSE} with $\beta=+\infty$. Therefore, one always
has, for $n$ large enough:
\begin{equation}
\label{situation1}
AMSE({\hat{\theta}}_n^
{(3)})<AMSE({\hat{\theta}}_n^{(1)})<AMSE({\hat{\theta}}_n^{(2)}).
\end{equation}

\paragraph{\bf Gamma distribution}$\Gamma(a,\lambda)$, $a, \lambda>0$.
We use the following parameterization of the density
$$
f(x)=\frac{\lambda^a}{\Gamma(a)} x^{a-1} \exp{(-\lambda x)}.
$$
From~\cite{EMBR}, Table 3.4.4, we obtain $\invH(x)= x^\theta \ell(x)$ with
$\theta=1$ and
$$
\ell(x)= \frac{1}{\lambda} + \frac{a-1}{\lambda} \frac{\log x}{x} + O(1/x).
$$
We thus have $\rho=-1$ and $b(x)=(1-a)\log(x)/x + O(1/x)$.
If $a>1$, $b$ is ultimately negative, corresponding to case i) of
Theorem~\ref{theoAMSE}. The conclusion depends on the value of $k_n$
as explained in the preceding section.  
If $a<1$, $b$ is ultimately positive, corresponding to case ii) of
Theorem~\ref{theoAMSE} with $\beta=+\infty$. 
Therefore, we are in situation~(\ref{situation1}).

\paragraph{\bf Weibull distribution}${\mathcal W}(a,\lambda)$, $a,\lambda>0$.
The inverse failure rate function is $\invH(x)= \lambda x^{1/a}$,
and then $\theta=1/a$, $\ell(x)=\lambda$ for all $x>0$. 
Therefore $b(x)=0$ and we
use the usual convention $\rho=-\infty$.
One may apply either i) or ii) of Theorem~\ref{theoAMSE}
with $\alpha=\beta=0$ to get
for $n$ large enough, 
\begin{equation}
\label{situation2}
AMSE({\hat{\theta}}_n^{(1)}) < \min(AMSE({\hat{\theta}}_n^{(2)}),AMSE({\hat{\theta}}_n^{(3)})).
\end{equation}

\subsection{Numerical results}

\begin{figure}[h]
\begin{center}
\epsfig{figure=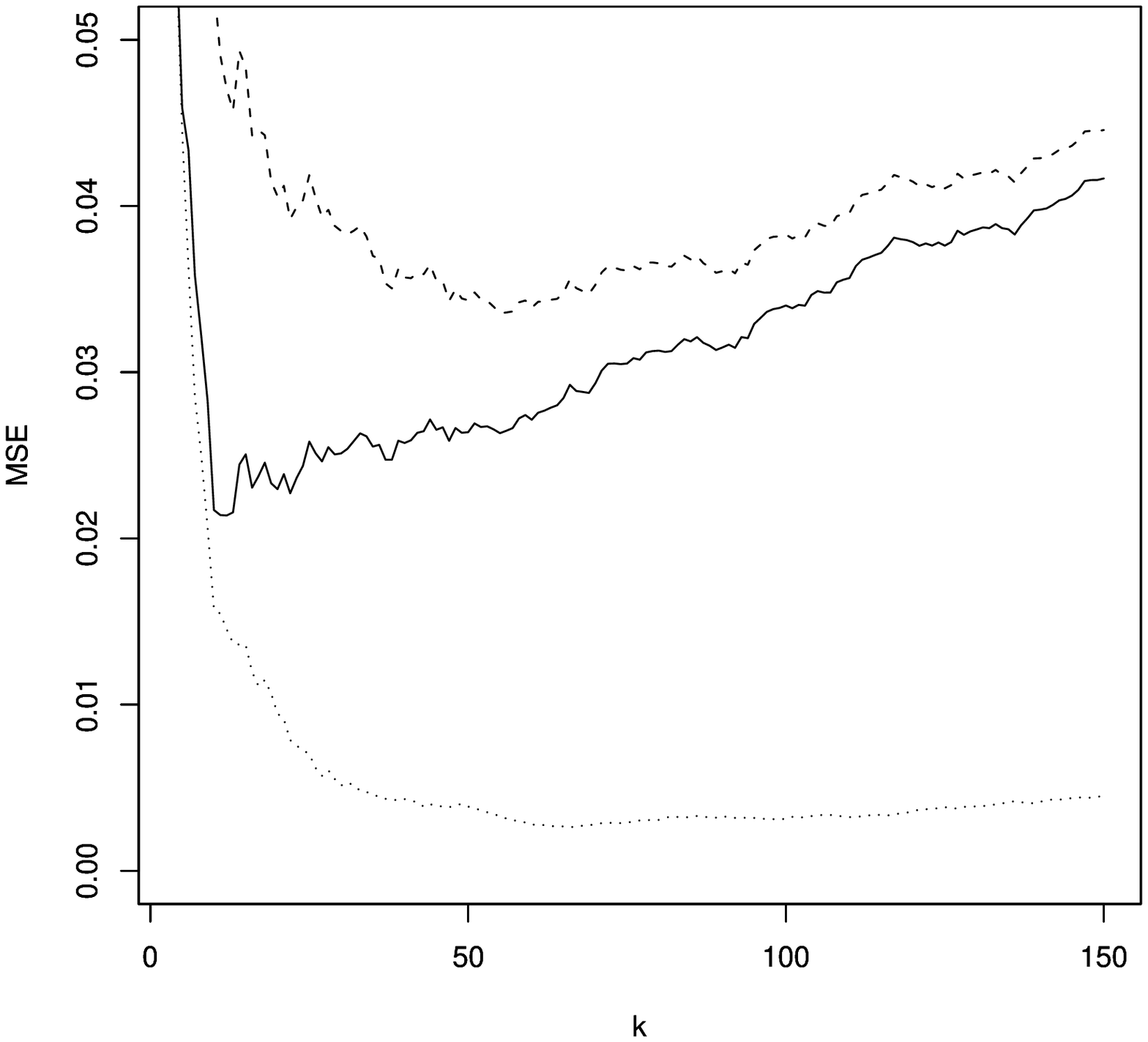,width=10cm,angle=0}

\epsfig{figure=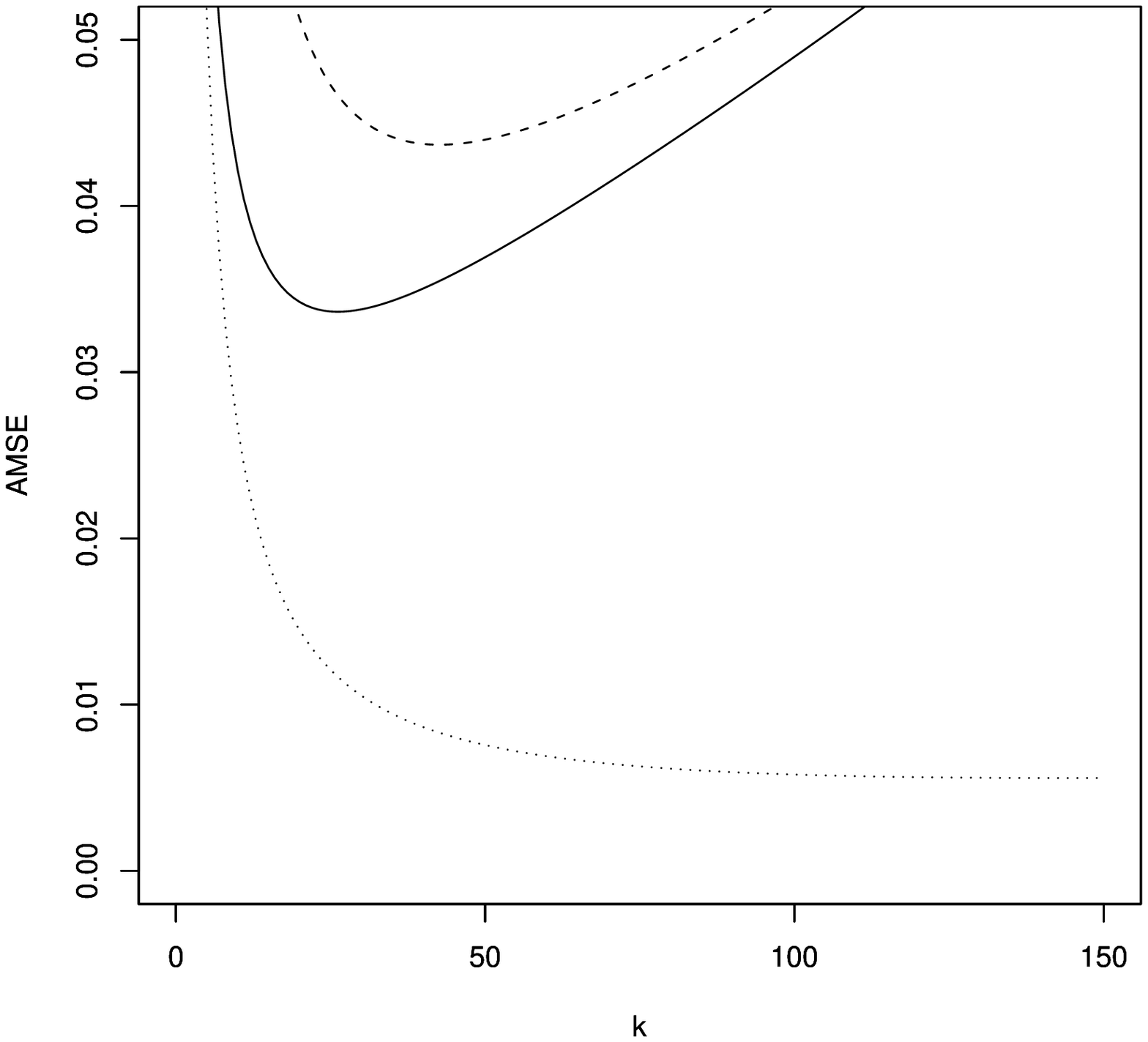,width=10cm,angle=0}
\end{center}
\caption{Comparison of estimates $\hat{\theta}_n^{(1)}$ (solid line), $\hat{\theta}_n^{(2)}$ (dashed line) and $\hat{\theta}_n^{(3)}$ (dotted line) for the $|{\cal{N}}(0,1)|$ distribution.  Up: MSE, down: AMSE. }
\label{figabsnor}
\end{figure}

\begin{figure}[h]
\begin{center}
\epsfig{figure=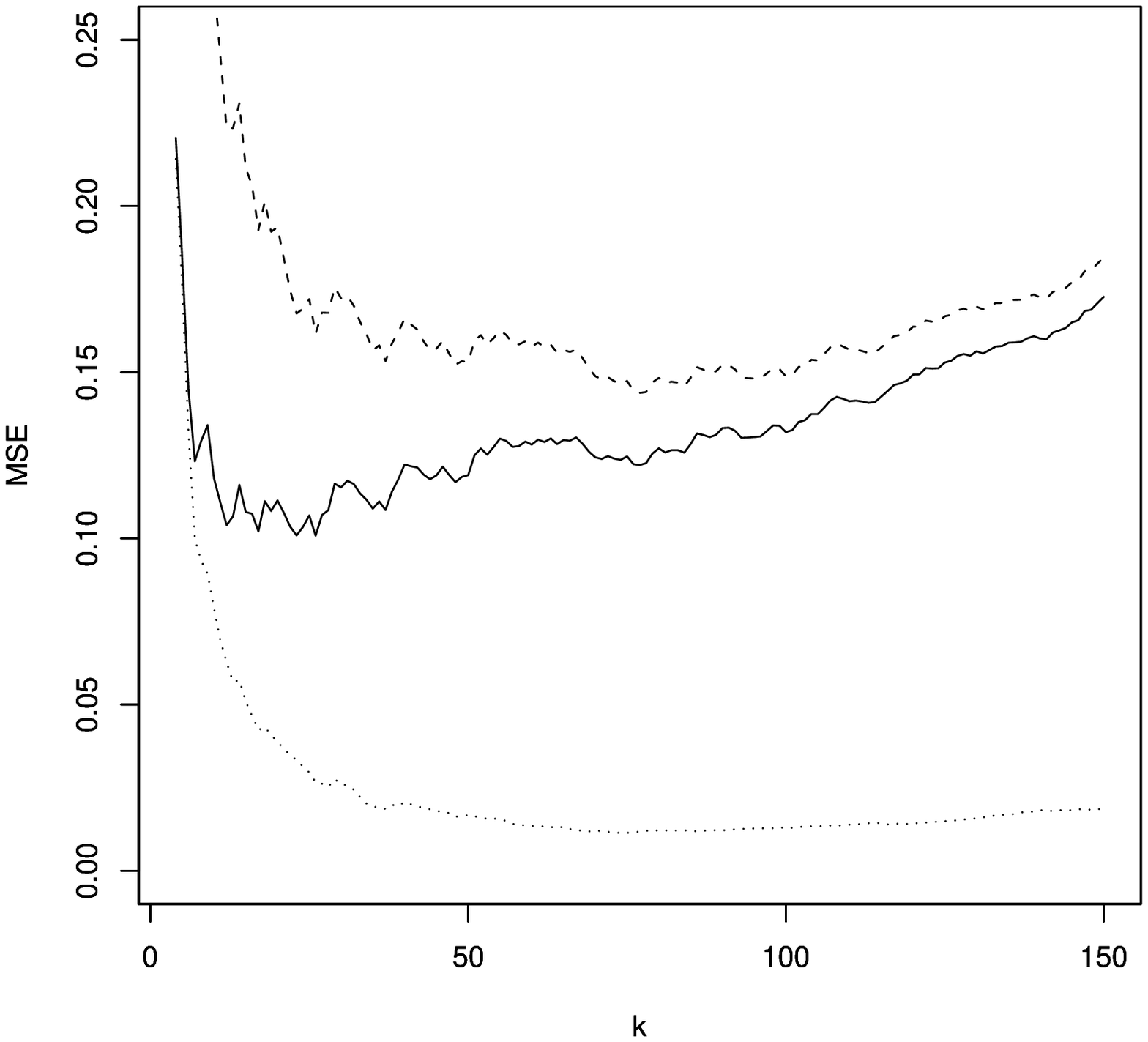,width=10cm,angle=0}

\epsfig{figure=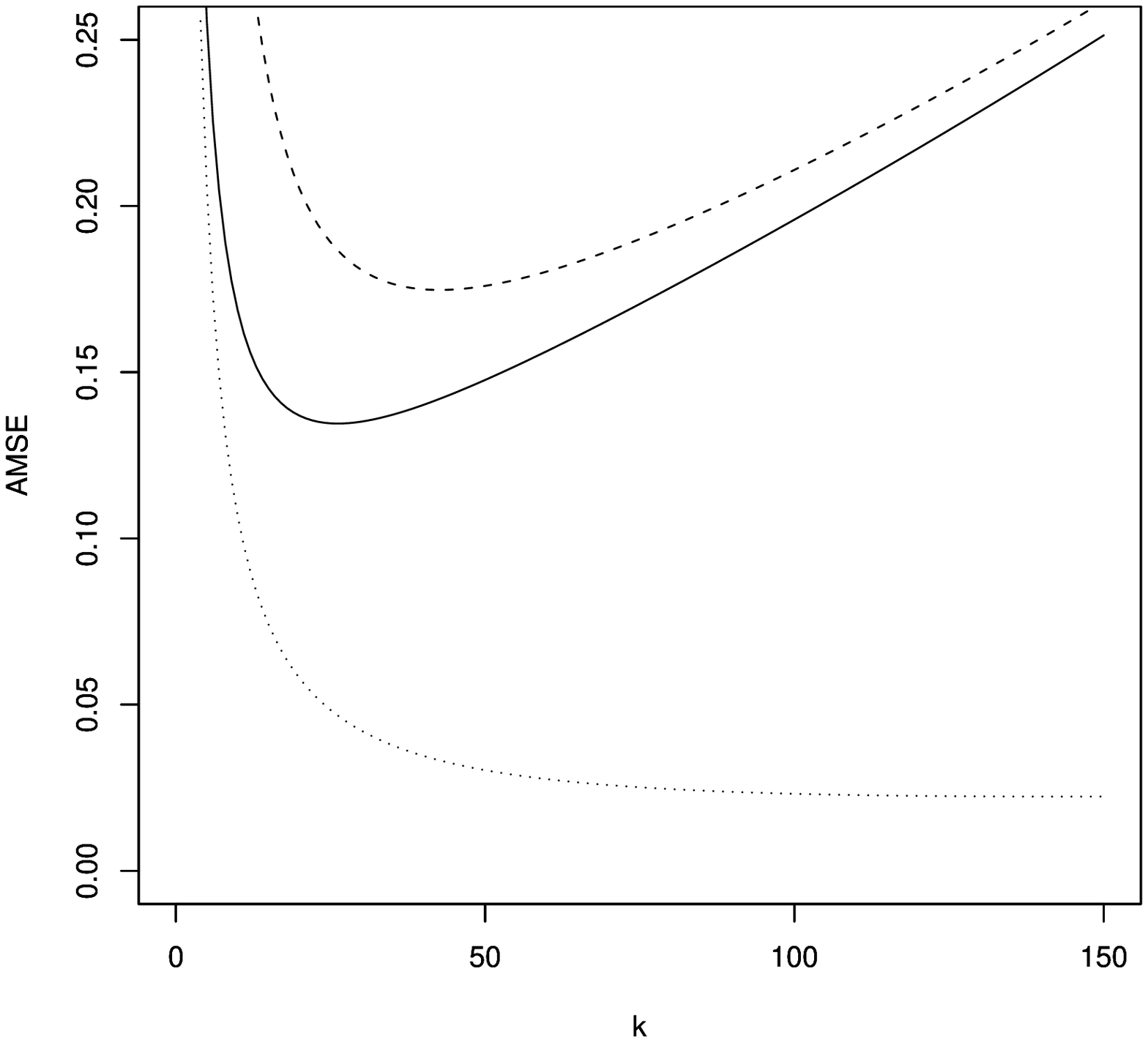,width=10cm,angle=0}
\end{center}
\caption{Comparison of estimates $\hat{\theta}_n^{(1)}$ (solid line), 
$\hat{\theta}_n^{(2)}$ (dashed line) and $\hat{\theta}_n^{(3)}$ (dotted line) 
for the $\Gamma(0.5,1)$ distribution.  Up: MSE, down: AMSE. }
\label{figgam05}
\end{figure}

\begin{figure}[h]
\begin{center}
\epsfig{figure=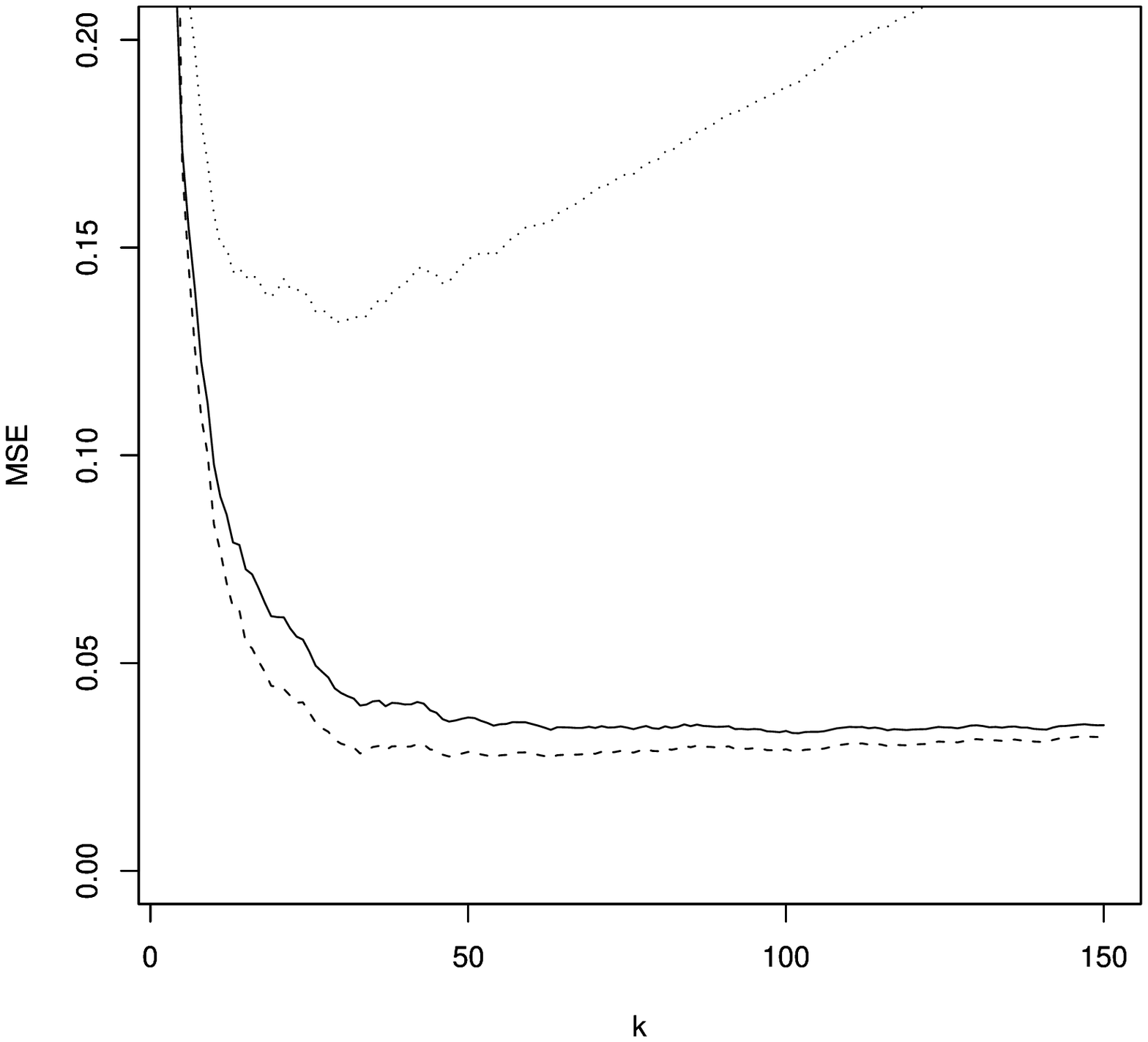,width=10cm,angle=0}

\epsfig{figure=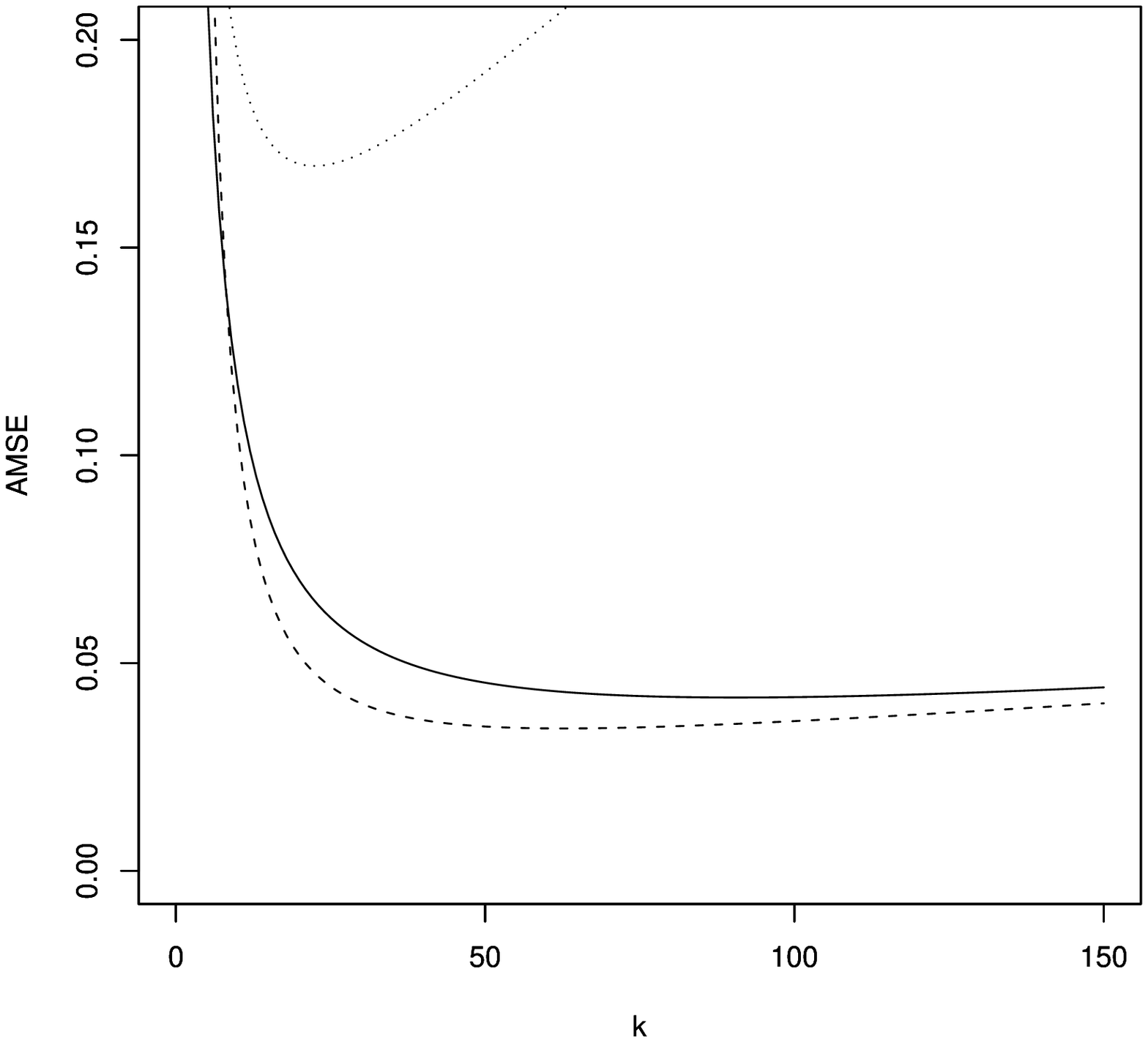,width=10cm,angle=0}
\end{center}
\caption{Comparison of estimates $\hat{\theta}_n^{(1)}$ (solid line), $\hat{\theta}_n^{(2)}$ (dashed line) and $\hat{\theta}_n^{(3)}$ (dotted line) for the $\Gamma(1.5,1)$ distribution.  Up: MSE, down: AMSE. }
\label{figgam15}
\end{figure}

\begin{figure}[h]
\begin{center}
\epsfig{figure=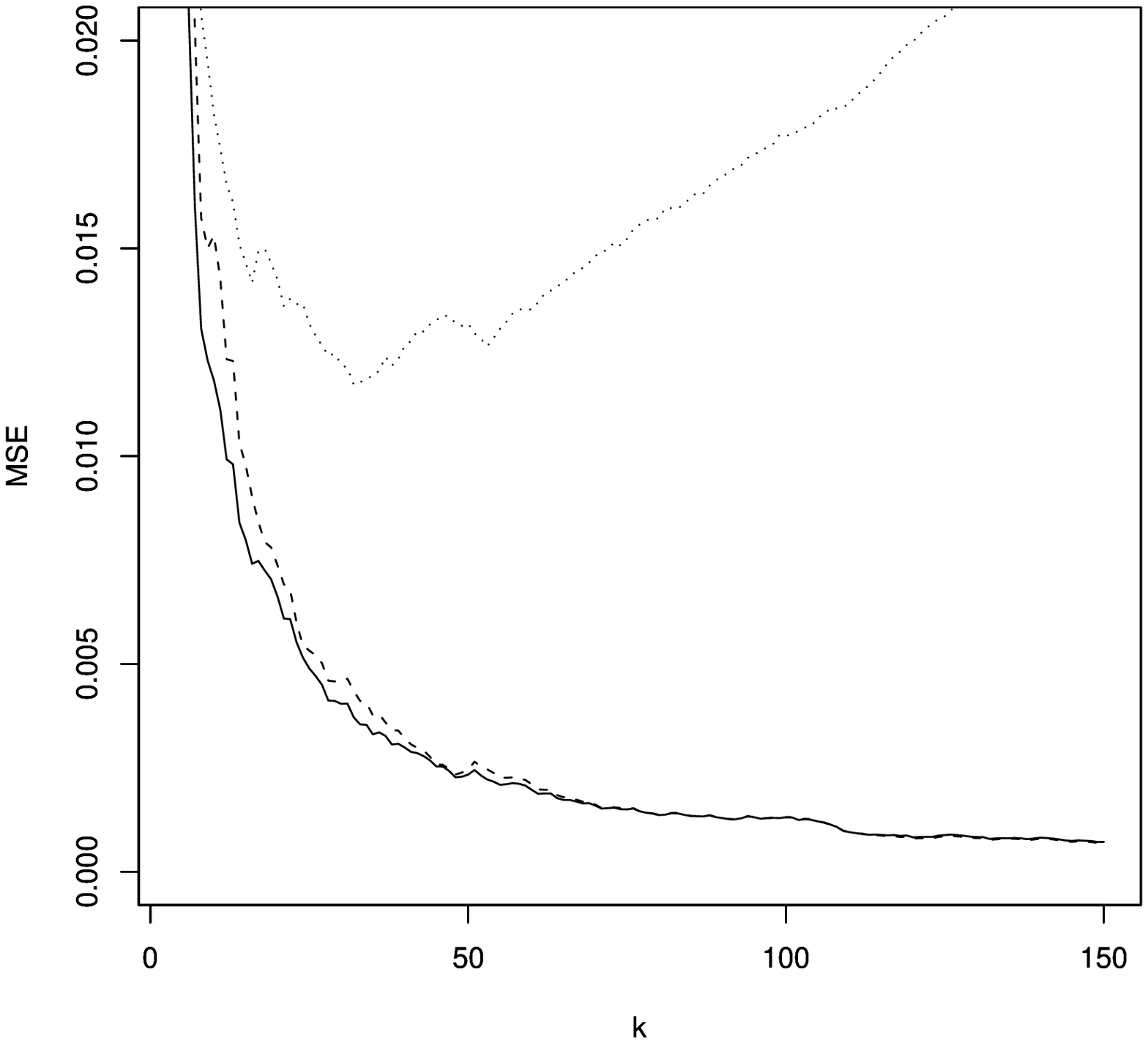,width=10cm,angle=0}

\epsfig{figure=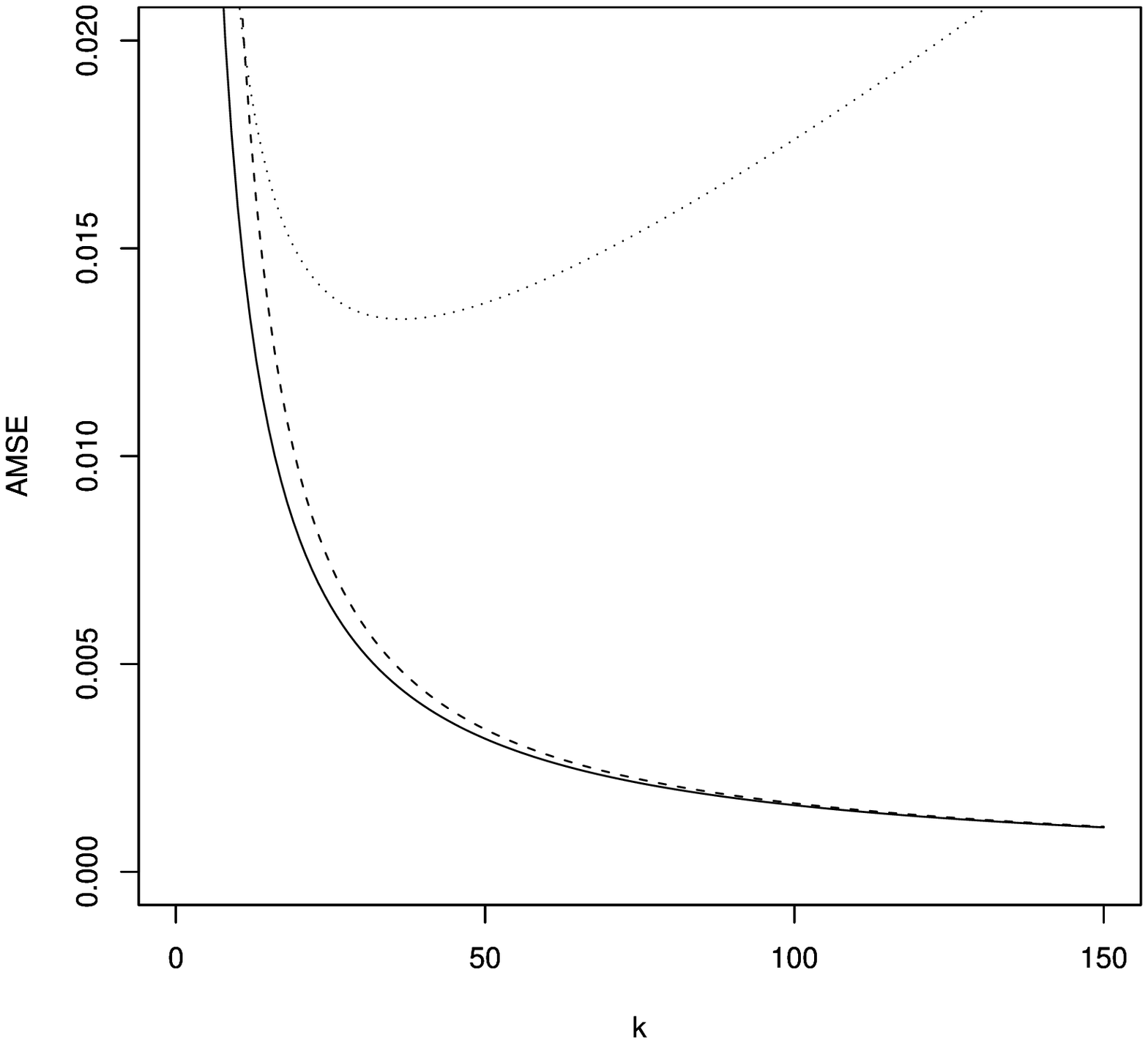,width=10cm,angle=0}
\end{center}
\caption{Comparison of estimates $\hat{\theta}_n^{(1)}$ (solid line), 
$\hat{\theta}_n^{(2)}$ (dashed line) and $\hat{\theta}_n^{(3)}$ (dotted line) 
for the ${\cal{W}}(2.5,2.5)$ distribution.  Up: MSE, down: AMSE.  }
\label{figwei25}
\end{figure}

\begin{figure}[h]
\begin{center}
\epsfig{figure=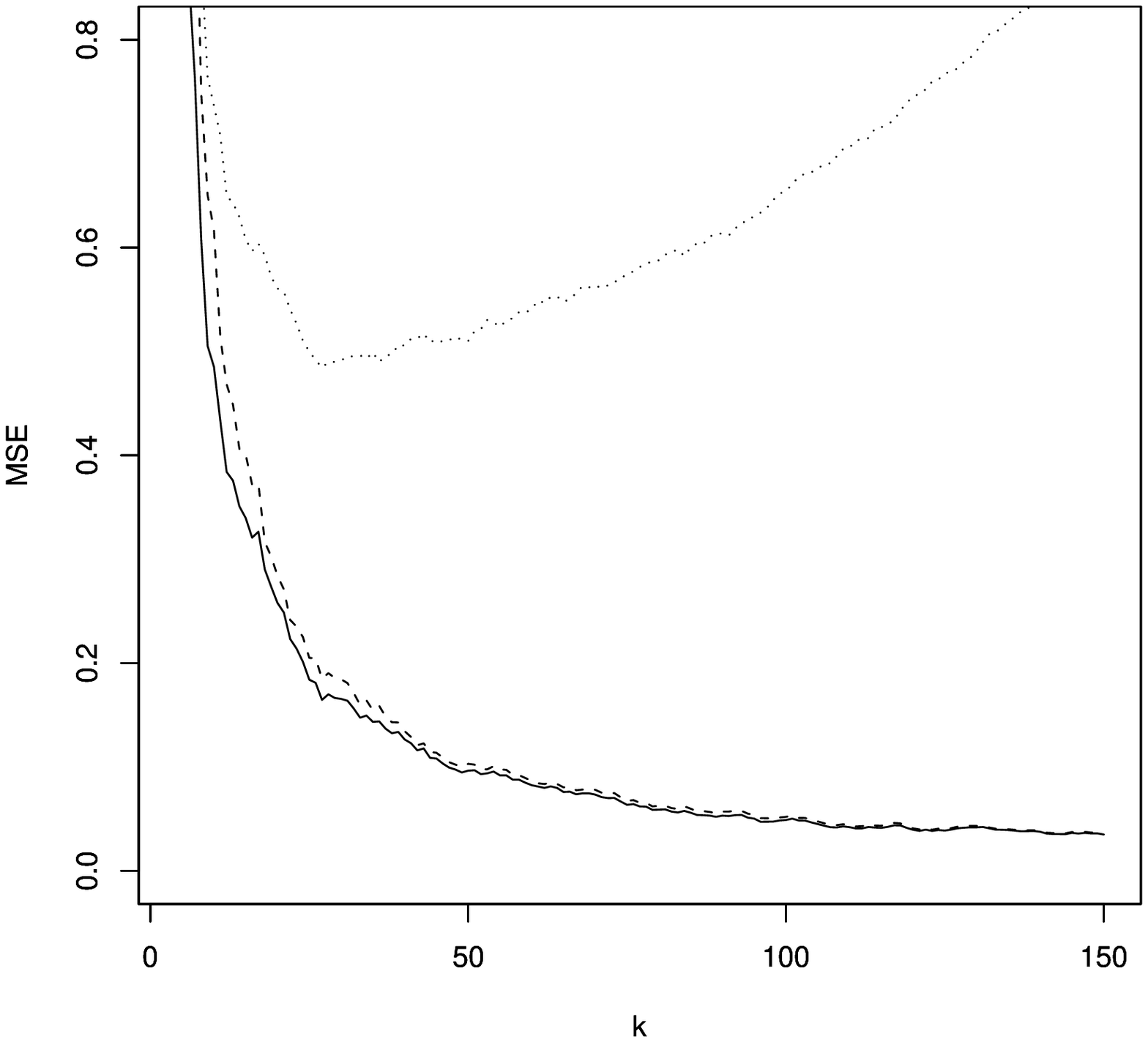,width=10cm,angle=0}

\epsfig{figure=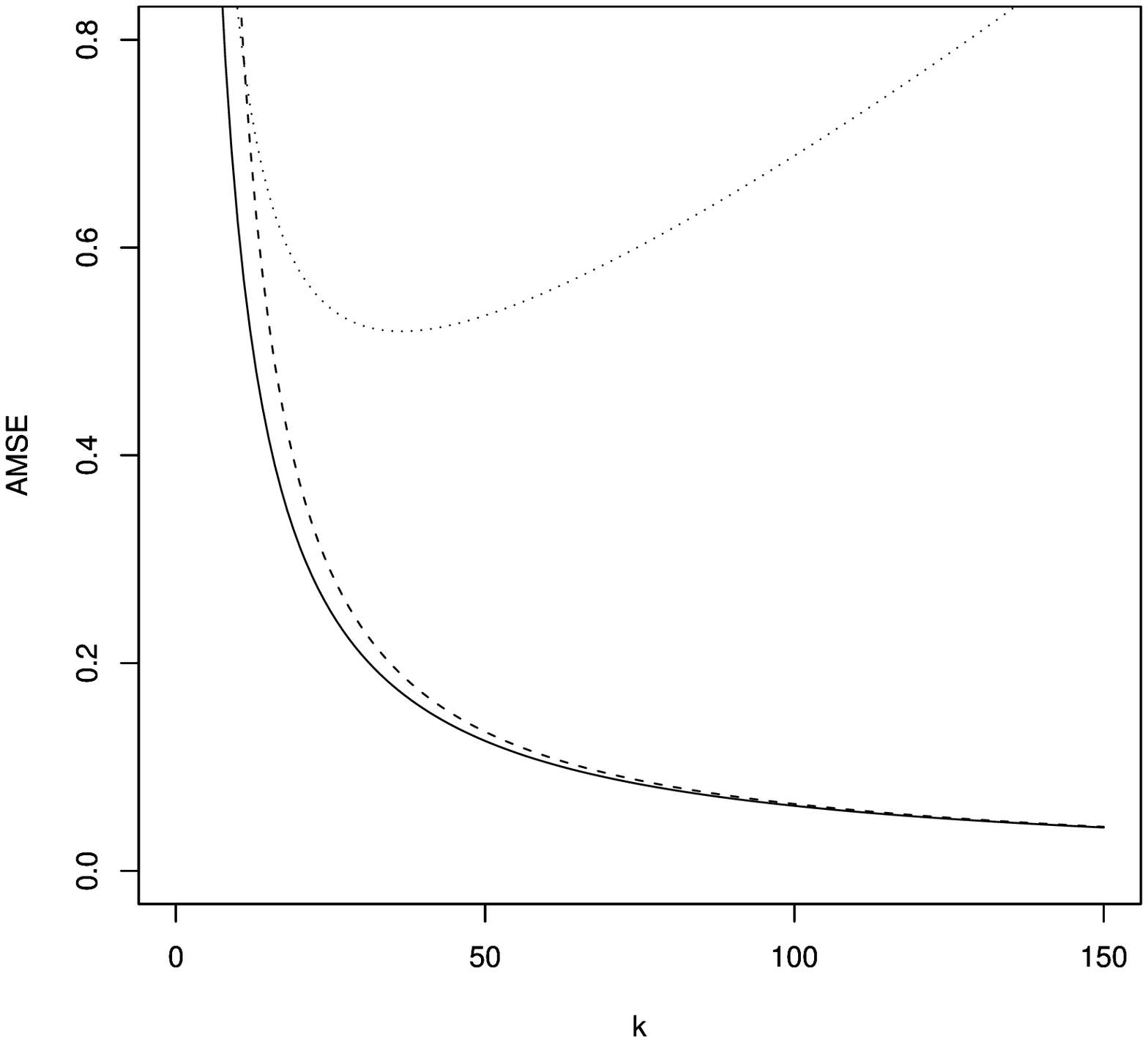,width=10cm,angle=0}
\end{center}
\caption{Comparison of estimates $\hat{\theta}_n^{(1)}$ (solid line), 
$\hat{\theta}_n^{(2)}$ (dashed line) and $\hat{\theta}_n^{(3)}$ (dotted line) 
for the ${\cal{W}}(0.4,0.4)$ distribution. Up: MSE, down: AMSE.  }
\label{figwei04}
\end{figure}

The finite sample performance of the estimators $\hat{\theta}_n^{(1)}$, $\hat{\theta}_n^{(2)}$ and $\hat{\theta}_n^{(3)}$ are investigated on~5 different distributions: $\Gamma(0.5,1)$, $\Gamma(1.5,1)$, $|{\mathcal N}(0,1)|$, ${\mathcal W}(2.5,2.5)$ and ${\mathcal W}(0.4,0.4)$.
In each case, $N=200$ samples $({\mathcal X}_{n,i})_{i=1,\dots,N}$
of size $n=500$ were simulated.
On each sample $({\mathcal X}_{n,i})$, the estimates $\hat{\theta}_{n,i}^{(1)}(k)$, $\hat{\theta}_{n,i}^{(2)}(k)$ and $\hat{\theta}_{n,i}^{(3)}(k)$ are computed for $k=2,\dots,150$.
Finally, the associated Mean Square Error (MSE) plots
are built by plotting the points
$$
\left( k, \frac{1}{N}\sum_{i=1}^N \left(\hat{\theta}_{n,i}^{(j)}(k)-\theta\right)^2\right) \ j=1, \ 2, \ 3.
$$
They are compared to the AMSE plots (see (\ref{AMSE}) for the definition of the AMSE):
$$
\left( k, (\theta a_n^{(j)} + b(\log(n/k)) )^2 + \frac{\theta^2}{k}\right) \ j=1, \ 2, \ 3,
$$
and where $b$ is given by (\ref{defb}). It appears on
Figure~\ref{figabsnor} -- Figure~\ref{figwei04} that,
for all the above mentioned distributions, the MSE and
AMSE have a similar qualitative behavior.
Figure~\ref{figabsnor} and Figure~\ref{figgam05} illustrate
situation~(\ref{situation1}) corresponding to ultimately positive
bias functions.  
The case of an ultimately negative bias function is presented
on Figure~\ref{figgam15} with the $\Gamma(1.5,1)$ distribution.  
It clearly appears that the MSE associated to $\thetachap^{(3)}$
is the largest.  For small values of $k$, one has
$MSE({\hat{\theta}}_n^{(1)}) < MSE({\hat{\theta}}_n^{(2)})$
and $MSE({\hat{\theta}}_n^{(1)}) > MSE({\hat{\theta}}_n^{(2)})$
for large value of $k$. This phenomenon is the illustration
of the asymptotic result presented in Theorem~\ref{theoAMSE}i).
Finally, Figure~\ref{figwei25} and Figure~\ref{figwei04} illustrate
situation~(\ref{situation2}) of asymptotically null bias functions.  
Note that, the MSE of $\thetachap^{(1)}$ and $\thetachap^{(2)}$
are very similar.  
As a conclusion, it appears that, in all situations,
$\thetachap^{(1)}$ and $\thetachap^{(2)}$ share a similar
behavior, with a small advantage to $\thetachap^{(1)}$ .
They provide good results for null and negative bias functions.  
At the opposite, $\thetachap^{(3)}$ should be preferred for
positive bias functions.

\section{Proofs}
\label{proofs}

For the sake of simplicity, in the following, we note $k$ for $k_n$.
We first give some preliminary lemmas.  Their proofs are postponed to the appendix.

\subsection{Preliminary lemmas}

We first quote a technical lemma.
\begin{Lem}
\label{new}
Suppose that $b$ is ultimately non-zero.  
If $k\to\infty$, $k/n \to 0$ and $k^{1/2}b(\log(n/k))\to\lambda\in\R$, then 
$\log(k)/\log(n)\to 0$.  
\end{Lem}
The following two lemmas are of analytical nature. They provide first-order expansions 
which will reveal useful in the sequel. 

\begin{Lem}
\label{lemmoments}
For all $\rho\leq 0$ and $q \in {\mathbb{N}}^*$, we have 
$$
\int_0^{\infty} K_{\rho}^q \left ( 1+\frac{x}{t}\right ) {\rm e}^{-x} dx \sim \frac{q!}{t^q} \ \mbox{as} \ t \to \infty.  $$
\end{Lem}
Let $a_n^{(i)}=\mu_{0}(\log(n/k_n))/T_n^{(i)}-1$, for $i=1,2,3$.  
\begin{Lem}
\label{an2an3}
Suppose  $k \to \infty$ and $k/n\to 0$.  
\begin{itemize}
\item[i)]  $T_n^{(1)} \log(n/k)\to 1$ and $a_n^{(1)}=0$.  
\item[ii)] $T_n^{(2)} \log(n/k)\to 1$. If moreover $\log(k)/\log(n)\to 0$ then $a_n^{(2)} \sim \log(k)/(2k)$.
\item[iii)] $T_n^{(3)} \log(n/k)= 1$ and $a_n^{(3)} \sim -1/\log(n/k)$.
\end{itemize}
\end{Lem}
The next lemma presents an expansion of $\hat{\theta}_n$. 
\begin{Lem}
\label{lemdecompo}
Suppose $k\to \infty$ and $k/n \to 0$. Under~\Au~and \Ad, the following expansions hold:
\[
 \hat{\theta}_n = \frac{1}{T_n} \left( \theta U_n^{(0)} + b(\log(n/k))U_n^{(\rho)} (1+o_{\rm{P}}(1)) \right ), 
\]
where 
\[ U_n^{(\rho)} = \frac{1}{k} \sum_{i=1}^{k-1} K_{\rho} \left ( 1+\frac{F_i}{E_{n-k+1,n}}\right ), \ \rho \leq 0 \]
and where $E_{n-k+1,n}$ is the $(n-k+1)$th order statistics associated to $n$ independent standard exponential variables and $\{F_1, \ldots, F_{k-1}\}$ are independent standard exponential variables and independent from $E_{n-k+1,n}$.
\end{Lem}
The next two lemmas provide the key results for establishing the asymptotic distribution of $\hat\theta_n$. 
Their describe they asymptotic behavior of the random terms appearing
in Lemma~\ref{lemdecompo}.

\begin{Lem}
\label{lemmoyvar}
Suppose $k \to \infty$ and $k/n \to 0$. Then, for all $\rho \leq 0$,
\[ \mu_{\rho}(E_{n-k+1,n}) \simP \sigma_{\rho}(E_{n-k+1,n}) \simP \frac{1}{\log(n/k)}. \]
\end{Lem}
\begin{Lem}
\label{lemNAUrho}
Suppose $k \to \infty$ and $k/n \to 0$. Then, for all $\rho \leq 0$,
\[ \frac{k^{1/2}}{\sigma_{\rho}(E_{n-k+1,n})} (U_n^{(\rho)}-\mu_{\rho}(E_{n-k+1,n})) \tod \mathcal{N}(0,1). \]
\end{Lem}

\subsection{Proofs of the main results}

\noindent {\bf{Proof of Proposition \ref{decomptheta}}} $-$ Lemma \ref{lemNAUrho} states that for $\rho \leq 0$,
\[ \frac{k^{1/2}}{\sigma_{\rho}(E_{n-k+1,n})} (U_n^{(\rho)}-\mu_{\rho}(E_{n-k+1,n}))=\xi_n(\rho), \]
where $\xi_n(\rho) \tod \mathcal{N}(0,1)$ for $\rho \leq 0$. Then, by Lemma~\ref{lemdecompo}
\begin{eqnarray*}
k^{1/2}(\hat{\theta}_n-\theta) &=& \theta \frac{\sigma_0(E_{n-k+1,n})}{T_n} \xi_n(0) + k^{1/2} \theta \left ( \frac{\mu_0(E_{n-k+1,n})}{T_n} -1\right) \\
 \ &+& k^{1/2}b(\log(n/k)) \left ( \frac{\sigma_{\rho}(E_{n-k+1,n})}{T_n} \frac{\xi_n(\rho)}{k^{1/2}} + \frac{\mu_{\rho}(E_{n-k+1,n})}{T_n}\right )(1+o_{{\rm{P}}}(1)).
\end{eqnarray*}
Since $T_n \sim 1/\log(n/k)$ and from Lemma~\ref{lemmoyvar}, we have
\begin{equation}
\label{eqtmp}
 k^{1/2}(\hat{\theta}_n -\theta) = \theta \xi_{n,1} +k^{1/2} \theta \left ( \frac{\mu_0(E_{n-k+1,n})}{T_n} -1\right) +k^{1/2}b(\log(n/k))(1+o_{\rm{P}}(1)), 
\end{equation}
where $\xi_{n,1} \tod \mathcal{N}(0,1)$. Moreover, a first-order expansion of $\mu_0$ yields
$$
 \frac{\mu_0(E_{n-k+1,n})}{\mu_0(\log(n/k))} = 1+(E_{n-k+1,n}-\log(n/k)) \frac{\mu_0^{(1)}(\eta_n)}{\mu_0(\log(n/k))},
$$
where $\eta_n \in ]\min(E_{n-k+1,n},\log(n/k)),\max(E_{n-k+1,n},\log(n/k))[$ and
\[ \mu_0^{(1)}(t) = \frac{d}{dt} \int_0^{\infty} \log \left ( 1+\frac{x}{t}\right ){\rm{e}}^{-x}dx =: \frac{d}{dt} \int_0^{\infty} f(x,t)dx. \]
Since for $t \geq T > 0$, $f(.,t)$ is integrable, continuous and
\[ \left | \frac{\partial f(x,t)}{\partial t} \right | = \frac{x}{t^2} \left ( 1+\frac{x}{t}\right )^{-1} {\rm{e}}^{-x} \leq x\frac{{\rm{e}}^{-x}}{T^2}, \]
we have that 
\[ \mu_0^{(1)}(t) = - \int_0^{\infty} \frac{x}{t^2} \left ( 1+\frac{x}{t}\right )^{-1} {\rm{e}}^{-x}dx. \]
Then, Lebesgue Theorem implies that $\mu_0^{(1)}(t) \sim -1/t^2$ as $t \to \infty$. 
Therefore, $\mu_0^{(1)}$ is regularly varying at infinity and thus
\[ \frac{\mu_0^{(1)}(\eta_n) }{\mu_0(\log(n/k))} \simP \frac{\mu_0^{(1)}(\log(n/k))}{\mu_0(\log(n/k))} \sim - \frac{1}{\log(n/k)}. \]
Since $k^{1/2}(E_{n-k+1,n}-\log(n/k)) \tod \mathcal{N} (0,1)$ (see~\cite{thetaWT}, Lemma 1), we have 
\begin{equation}
\label{mu0Emu0log}
\frac{\mu_0(E_{n-k+1,n})}{\mu_0(\log(n/k))} = 1 - \frac{k^{-1/2}}{\log(n/k)} \xi_{n,2}, 
\end{equation}
where $\xi_{n,2} \tod \mathcal{N} (0,1)$. Collecting (\ref{eqtmp}), (\ref{mu0Emu0log}) and
taking into account that $T_n\log(n/k)\to 1$ concludes the proof. \CQFD


\noindent {\bf{Proof of Proposition \ref{vitesse}}} $-$ Lemma~\ref{new} entails 
$\log(n/k) \sim \log(n)$. Since $|b|$ is a regularly varying function, $b(\log(n/k)) \sim b(\log(n))$ and thus, $k^{1/2} \sim \lambda/b(\log(n))$. \CQFD


\noindent {\bf{Proof of Theorem \ref{NAquant}}} $-$ The asymptotic
normality of $\qchap$ can be deduced from the asymptotic normality
of $\thetachap$ using Theorem~2.3 of~\cite{quantWT}. We are
in the situation, denoted by {\bf (S.2)} in the above mentioned paper,
where the limit distribution of $\qchap/\qvrai$ is driven by $\thetachap$.
Following, the notations of~\cite{quantWT}, we denote by
$\alpha_n=k_n^{1/2}$ the asymptotic rate of convergence of $\thetachap$,
by $\beta_n=\theta a_n$ its asymptotic bias, and by 
${\cal L}={\cal N}(0,\theta^2)$ its asymptotic distribution.  
It suffices to verify that 
\begin{equation}
\label{averifier}
\log(\tau_n)\log(n/k)\to\infty.  
\end{equation}
To this end, note that conditions~(\ref{condcons}) and $p_n<1/n$ imply
that there exists $0<c<1$ such that
$$
\log(\tau_n)> c (\tau_n-1) > c \left(\frac{\log(n)}{\log(n/k)}-1\right)
=c \frac{\log(k)}{\log(n/k)},
$$
which proves~(\ref{averifier}).
We thus have
\[
\frac{k^{1/2}}{\log \tau_n}\tau_n^{-\theta a_n}  \left(\frac{\qchap}{\qvrai}-\tau_n^{\theta a_n} \right) \tod {\mathcal{N}}(0,\theta^2). 
\]
Now, remarking that, from Lemma~\ref{lemmoments}, 
$\mu_0(\log(n/k))\sim 1/\log(n/k)\sim T_n$, and thus $a_n\to 0$
gives the result.  \CQFD

\noindent {\bf{Proof of Corollary \ref{CorNAtroistheta}}} $-$ 
Lemma~\ref{an2an3} shows that the assumptions of Theorem~\ref{NAtheta} 
and Theorem~\ref{NAquant} are
verified and that, for $i=1,2,3$, $k^{1/2} a_n^{(i)} \to 0$. \CQFD


\noindent {\bf{Proof of Theorem \ref{theoAMSE}}} $-$ \\
i) First, from~(\ref{AMSE}) and Lemma \ref{an2an3} iii), since $b$ is ultimately non-positive,
\begin{equation}
\label{A}
AMSE({\hat{\theta}}_n^{(1)})-AMSE({\hat{\theta}}_n^{(3)}) = -\theta (a_n^{(3)})^2 \left( \theta  +
2\frac{b(\log(n/k))}{a_n^{(3)}}\right)  < 0.
\end{equation}
Second, from~(\ref{AMSE}),
\begin{equation}
\label{B}
AMSE({\hat{\theta}}_n^{(2)})-AMSE({\hat{\theta}}_n^{(1)})=
\theta(a_n^{(2)})^2 \left( \theta  + 
2\frac{b(\log(n/k))}{a_n^{(2)}}\right). 
\end{equation}
If $b$ is ultimately non-zero, Lemma~\ref{new} entails that $\log(n/k) \sim \log(n)$ 
and consequently, since $|b|$ is regularly varying, 
$b(\log(n/k)) \sim b(\log(n))$.   
Thus, from Lemma \ref{an2an3} ii),
\begin{equation}
\label{C}
2\frac{b(\log(n/k))}{a_n^{(2)}} \sim 4 b(\log n)\frac{k}{\log(k)} \to -\alpha. 
\end{equation}
Collecting (\ref{A})--(\ref{C}) concludes the proof of i). 
\\
\noindent ii) First, (\ref{B}) and  Lemma \ref{an2an3} ii) yields
\begin{equation}
\label{D}
AMSE({\hat{\theta}}_n^{(2)})-AMSE({\hat{\theta}}_n^{(1)})  > 0,
\end{equation}
since $b$ is ultimately non-negative. Second, if $b$ is ultimately non-zero, Lemma~\ref{new} entails that $\log(n/k) \sim \log(n)$ 
and consequently, since $|b|$ is regularly varying, 
$b(\log(n/k)) \sim b(\log(n))$.   
Thus, observe that in (\ref{A}), 
\begin{equation}
\label{E}
2\frac{b(\log(n/k))}{a_n^{(3)}} \sim -2 b(\log n)(\log n) \to -\beta. 
\end{equation}
Collecting  (\ref{A}), (\ref{D}) and (\ref{E}) concludes the proof of ii).
The case when $b$ is ultimately zero is obtained either by considering $\alpha=0$ in~(\ref{C}),
or $\beta=0$ in (\ref{E}).
\CQFD

\newpage

\newpage
\section*{Appendix: proof of lemmas}

\noindent {\bf{Proof of Lemma~\ref{new}}} $-$ Remark that, for $n$ large enough,
$$
|k^{1/2}b(\log(n/k))| \leq |k^{1/2}b(\log(n/k))-\lambda| + |\lambda| \leq 1 + |\lambda|,
$$
and thus, if $b$ is ultimately non-zero,
\begin{equation}
\label{passagelog}
0\leq \frac{1}{2} \frac{\log(k)}{\log(n/k)} \leq  \frac{\log{(1+|\lambda|)}}{\log(n/k)} - 
\frac{\log |b(\log(n/k))|}{\log(n/k)}.
\end{equation}
Since $|b|$ is a regularly varying function, we have that (see~\cite{bingolteu87}, Proposition 1.3.6.)
\[ \frac{\log |b(\log(x))|}{\log(x)} \to 0 \ {\mbox{as}} \ x \to \infty. \]
Then, (\ref{passagelog}) implies $\log(k)/\log(n/k) \to 0$ which entails 
$\log(k)/\log(n)\to 0$.  \CQFD


\noindent {\bf{Proof of Lemma~\ref{lemmoments}}} $-$ 
Since for all $x,t >0$, $tK_{\rho}(1+x/t)<x$, Lebesgue Theorem implies that
\[ 
\lim_{t \to \infty} \int_0^\infty\left ( t K_{\rho} \left ( 1+\frac{x}{t} \right )\right )^q {\rm{e}}^{-x}dx =
\int_0^{\infty} \lim_{t \to \infty}\left ( t K_{\rho} \left ( 1+\frac{x}{t} \right )\right )^q {\rm{e}}^{-x}dx = \int_0^{\infty} x^q {\rm{e}}^{-x}dx = q!, \]
which concludes the proof.
\CQFD


\noindent {\bf{Proof of Lemma~\ref{an2an3}}} $-$ \\
\noindent i) Lemma~\ref{lemmoments} shows that $\mu_0(t)\sim 1/t$ 
and thus $T_n^{(1)}\log(n/k)\to 1$.  By definition, $a_n^{(1)}=0$.\\

\noindent ii) The well-known inequality $-x^2/2\leq\log (1+x) -x \leq 0$, $x>0$ yields
\begin{equation}
\label{encadr}
-\frac{1}{2} \frac{1}{\log(n/k)} \frac{1}{k} \sum_{i=1}^k \log^2(k/i)
\leq \log(n/k) T_n^{(2)} -\frac{1}{k} \sum_{i=1}^k \log(k/i) \leq 0.
\end{equation}
Now, since when $k\to\infty$,
$$
\frac{1}{k} \sum_{i=1}^k \log^2(k/i) \to \int_0^1 \log^2(x)dx =2
\mbox{ and }
\frac{1}{k} \sum_{i=1}^k \log(k/i) \to -\int_0^1 \log(x)dx =1,
$$
it follows that $T_n^{(2)}\log(n/k)\to 1$.  
Let us now introduce the function defined on $(0,1]$ by:
\[ f_n(x)=\log \left ( 1-\frac{\log(x)}{\log(n/k)} \right ). \]
We have:
\begin{eqnarray*}
a_n^{(2)} & = & -\frac{1}{T_n^{(2)}} ( T_n^{(2)}-\mu_0(\log(n/k))) = -\frac{1}{T_n^{(2)}} \left ( \frac{1}{k} \sum_{i=1}^{k-1} f_n(i/k)-\int_0^1f_n(t)dt \right ) \\
 \ & = & -\frac{1}{T_n^{(2)}} \sum_{i=1}^{k-1} \int_{i/k}^{(i+1)/k} (f_n(i/k)-f_n(t))dt + \frac{1}{T_n^{(2)}} \int_0^{1/k} f_n(t)dt.
\end{eqnarray*}
Since
\[ f_n(t)=f_n \left ( i/k \right )+\left ( t - i/k \right ) f_n^{(1)} \left ( i/k \right ) + \int_{i/k}^t (t-x)f_n^{(2)}(x)dx, \]
where $f_n^{(p)}$ is the $p$th derivative of $f_n$, we have:
\begin{eqnarray*}
a_n^{(2)} &=& \frac{1}{T_n^{(2)}} \sum_{i=1}^{k-1} \int_{i/k}^{(i+1)/k} (t-i/k)f_n^{(1)}(i/k)dt  \\
 \ &+& \frac{1}{T_n^{(2)}} \sum_{i=1}^{k-1} \int_{i/k}^{(i+1)/k} \int_{i/k}^t (t-x)f_n^{(2)}(x)dx dt + \frac{1}{T_n^{(2)}} \int_0^{1/k}f_n(t)dt =: \Psi_1+\Psi_2+\Psi_3.
\end{eqnarray*}
Let us focus first on the term $\Psi_1$:
\begin{eqnarray*}
\Psi_1 & = & \frac{1}{T_n^{(2)}}\frac{1}{2k^2}\sum_{i=1}^{k-1} f_n^{(1)}(i/k) \\
 \ & = &  \frac{1}{2kT_n^{(2)}} \int_{1/k}^1f_n^{(1)}(x)dx + \frac{1}{2kT_n^{(2)}} \left ( \frac{1}{k} \sum_{i=1}^{k-1} f_n^{(1)}(i/k) - \int_{1/k}^1f_n^{(1)}(x)dx \right ) \\
 \ & = & \frac{1}{2kT_n^{(2)}} (f_n(1)-f_n(1/k)) - \frac{1}{2kT_n^{(2)}}\sum_{i=1}^{k-1}\int_{i/k}^{(i+1)/k} (f_n^{(1)}(x)-f_n^{(1)}(i/k))dx =: \Psi_{1,1}-\Psi_{1,2}.
\end{eqnarray*}
Since $T_n^{(2)} \sim 1/\log(n/k)$ and $\log(k)/\log(n) \to 0$, we have:
\[ \Psi_{1,1}  = -\frac{1}{2kT_n^{(2)}} \log \left ( 1+\frac{\log(k)}{\log(n/k)} \right ) = -\frac{\log(k)}{2k} (1+o(1)). \]
Furthermore, since, for $n$ large enough, $f_n^{(2)}(x) > 0$ for $x \in [0,1]$,
\begin{eqnarray*}
O\leq \Psi_{1,2} & \leq & \frac{1}{2kT_n^{(2)}}\sum_{i=1}^{k-1}\int_{i/k}^{(i+1)/k} (f_n^{(1)}((i+1)/k)-f_n^{(1)}(i/k))dx = \frac{1}{2k^2T_n^{(2)}} (f_n^{(1)}(1)-f_n^{(1)}(1/k)) \\
 \ & = & \frac{1}{2k^2T_n^{(2)}}\left ( -\frac{1}{\log(n/k)} + \frac{k}{\log(n/k)} \left ( 1+\frac{\log(k)}{\log(n/k)} \right )^{-1} \right ) \sim \frac{1}{2k} = o \left ( \frac{\log(k)}{k} \right ).
\end{eqnarray*}
Thus,
\begin{equation}
\label{T1}
\Psi_1 =-\frac{\log(k)}{2k} (1+o(1)).
\end{equation}
Second, let us focus on the term $\Psi_2$. Since, for $n$ large enough, $f_n^{(2)}(x) > 0$ for $x \in [0,1]$,
\begin{eqnarray}
\label{T2}
0\leq\Psi_2 & \leq & \frac{1}{T_n^{(2)}} \sum_{i=1}^{k-1}\int_{i/k}^{(i+1)/k} \int_{i/k}^{(i+1)/k} (t-i/k)f_n^{(2)}(x) dx dt \nonumber \\
 \ & = &\frac{1}{2k^2T_n^{(2)}}(f_n^{(1)}(1)-f_n^{(1)}(1/k)) =o \left ( \frac{\log(k)}{k} \right ).
\end{eqnarray}
Finally,
\[ \Psi_3 = \frac{1}{T_n^{(2)}} \int_0^{1/k} - \frac{\log(t)}{\log(n/k)} dt + \frac{1}{T_n^{(2)}} \int_0^{1/k} \left ( f_n(t) +\frac{\log(t)}{\log(n/k)} \right )dt =: \Psi_{3,1} + \Psi_{3,2}, \]
and we have:
\[ \Psi_{3,1} = \frac{1}{\log(n/k) T_n^{(2)}} \frac{1}{k} (\log(k)+1) = \frac{\log(k)}{k}(1+o(1)). \]
Furthermore, using the well known inequality: $|\log(1+x)-x| \leq x^2/2$, $x>0$, we have:
\begin{eqnarray*}
|\Psi_{3,2}| & \leq & \frac{1}{2T_n^{(2)}} \int_0^{1/k} \left ( \frac{\log(t)}{\log(n/k)} \right )^2 dt = \frac{1}{2T_n^{(2)}} \frac{1}{k(\log(n/k))^2} ((\log(k))^2+2\log(k)+2) \\
 \ & \sim & \frac{(\log(k))^2}{2k\log(n/k)} = o \left ( \frac{\log(k)}{k} \right ),
\end{eqnarray*}
since $\log(k)/\log(n) \to 0$. Thus,
\begin{equation}
\label{T3}
\Psi_3=\frac{\log(k)}{k}(1+o(1)).
\end{equation}
We conclude the proof of i) by collecting (\ref{T1})-(\ref{T3}). \\
\noindent ii) 
First, $T_n^{(3)}\log(n/k)=1$ by definition. Besides, we have 
\begin{eqnarray*}
a_n^{(3)} & = & \frac{\mu_0(\log(n/k))}{T_n^{(3)}} - 1 = \log(n/k) \mu_0(\log(n/k)) - 1 \\
 \ & = & \int_0^{\infty} \log(n/k) \log \left ( 1+\frac{x}{\log(n/k)} \right ) {\rm{e}}^{-x} dx - 1 \\
 \ & = & \int_0^{\infty} x {\rm{e}}^{-x} dx - \frac{1}{2} \int_0^{\infty} \frac{x^2}{\log(n/k)} {\rm{e}}^{-x} dx - 1 + R_n = -\frac{1}{\log(n/k)}+R_n,
\end{eqnarray*}
where
\[ R_n = \int_0^{\infty} \log(n/k) \left ( \log \left ( 1+\frac{x}{\log(n/k)} \right ) - \frac{x}{\log(n/k)} + \frac{x^2}{2(\log(n/k))^2} \right ) {\rm{e}}^{-x} dx. \]
Using the well known inequality: $|\log(1+x)-x+x^2/2| \leq x^3/3$, $x>0$, we have,
\[ |R_n| \leq \frac{1}{3}\int_0^{\infty}\frac{x^3}{(\log(n/k))^2}{\rm{e}}^{-x} dx =o \left ( \frac{1}{\log(n/k)} \right ), \]
which finally yields $a_n^{(3)}\sim -1/\log(n/k)$.
\CQFD


\noindent {\bf{Proof of Lemma~\ref{lemdecompo}}} $-$ Recall that 
$$
{\hat{\theta}}_n =: \frac{1}{T_n}\frac{1}{k} \somi  (\log(X_{n-i+1,n})  - \log(X_{n-k+1,n})),
$$
and let $E_{1,n},\dots,E_{n,n}$ be ordered statistics generated
by $n$ independent standard exponential random variables.
Under \Au, we have
\begin{eqnarray*}
{\hat{\theta}}_n & \egl & \frac{1}{T_n} \frac{1}{k} \somi (\log\invH(E_{n-i+1,n}) - \log\invH(E_{n-k+1,n})) \\
& \egl & \frac{1}{T_n} \left ( \theta\frac{1}{k} \somi \log\left(\frac{E_{n-i+1,n}}{E_{n-k+1,n}} \right)
+ \frac{1}{k} \somi \log\left(\frac{\ell(E_{n-i+1,n})}{\ell(E_{n-k+1,n})} \right) \right ).
\end{eqnarray*}
Define $x_n=E_{n-k+1,n}$ and $\lambda_{i,n}=E_{n-i+1,n}/E_{n-k+1,n}$.
It is clear, in view of~\cite{thetaWT}, Lemma~1 that $x_n\toP \infty$ and
$\lambda_{i,n}\toP 1$. Thus, \Ad~yields that 
uniformly in $i=1,\dots,k-1$:
$$
{\hat{\theta}}_n \egl \frac{1}{T_n} \left ( \theta\frac{1}{k}\somi \log\left(\frac{E_{n-i+1,n}}{E_{n-k+1,n}} \right) + (1+o_p(1)) b(E_{n-k+1,n}) \frac{1}{k}\somi K_\rho\left(\frac{E_{n-i+1,n}}{E_{n-k+1,n}} \right) \right ).
$$
The R\'enyi representation of the Exp(1) ordered statistics (see \cite{order}, p. 72)
yields
\begin{equation}
\label{repres}
\left\{ \frac{E_{n-i+1,n}}{E_{n-k+1,n}} \right\}_{i=1,\dots,k-1} 
\egl \left\{ 1+ \frac{F_{k-i,k-1}}{E_{n-k+1,n}} \right\}_{i=1,\dots,k-1},
\end{equation}
where $\{F_{1,k-1},\dots,F_{k-1,k-1}\}$ are ordered statistics
independent from $E_{n-k+1,n}$ and
generated by $k-1$ independent standard exponential variables 
$\{F_1,\dots,F_{k-1}\}$.
Therefore,
\begin{eqnarray*}
{\hat{\theta}}_n &\egl& \frac{1}{T_n} \left ( \theta\frac{1}{k} \somi \log\left(1+\frac{F_i}{E_{n-k+1,n}} \right) \right .\\
&+& \left . (1+o_p(1)) b(E_{n-k+1,n}) \frac{1}{k}\somi K_\rho\left(1+\frac{F_i}{E_{n-k+1,n}}\right) \right ).
\end{eqnarray*}
Remarking that 
$K_0(x)=\log(x)$ concludes the proof.
\CQFD


\noindent {\bf{Proof of Lemma~\ref{lemmoyvar}}} $-$ 
Lemma~\ref{lemmoments} implies that,
\[ \mu_{\rho}(E_{n-k+1,n}) \simP \frac{1}{E_{n-k+1,n}} \simP \frac{1}{\log(n/k)}, \]
since $E_{n-k+1,n}/\log(n/k) \toP 1$ (see \cite{thetaWT}, Lemma 1). Next, from Lemma~\ref{lemmoments},
\begin{eqnarray*}
 \sigma_{\rho}^2 (E_{n-k+1,n}) & = & \frac{2}{E_{n-k+1,n}^2}(1+o_{\rm{P}}(1))-\frac{1}{E_{n-k+1,n}^2}(1+o_{\rm{P}}(1)) \\
  \ & = & \frac{1}{E_{n-k+1,n}^2}(1+o_{\rm{P}}(1)) = \frac{1}{(\log(n/k))^2}(1+o_{\rm{P}}(1)),
\end{eqnarray*}
which concludes the proof.
\CQFD

\noindent {\bf{Proof of Lemma~\ref{lemNAUrho}}} $-$ Remark that
\begin{eqnarray*}
\frac{k^{1/2}}{\sigma_{\rho}(E_{n-k+1,n})} \left ( U_n^{(\rho)} -\mu_{\rho}(E_{n-k+1,n}) \right ) & = & \frac{k^{-1/2}}{\sigma_{\rho}(E_{n-k+1,n})} \sum_{i=1}^{k-1} \left ( K_{\rho} \left ( 1+\frac{F_i}{E_{n-k+1,n}} \right ) -\mu_{\rho}(E_{n-k+1,n}) \right ) \\
 \ & - & k^{-1/2} \frac{\mu_{\rho}(E_{n-k+1,n})}{\sigma_{\rho}(E_{n-k+1,n})}.
\end{eqnarray*}
Let us introduce the following notation:
\[ S_n(t)=\frac{(k-1)^{-1/2}}{\sigma_{\rho}(t)} \sum_{i=1}^{k-1} \left ( K_{\rho} \left ( 1+\frac{F_i}{t} \right ) - \mu_{\rho}(t) \right ). \]
Thus,
\[ \frac{k^{1/2}}{\sigma_{\rho}(E_{n-k+1,n})} \left ( U_n^{(\rho)} -\mu_{\rho}(E_{n-k+1,n}) \right ) = S_n(E_{n-k+1,n})(1+o(1)) + o_{{\rm{P}}}(1), \]
from Lemma~\ref{lemmoyvar}. It remains to prove that for $x \in \R$,
\[ {\rm{P}}(S_n(E_{n-k+1,n}) \leq x ) - \Phi(x) \to 0 \ {\mbox{as}} \ n \to \infty, \]
where $\Phi$ is the cumulative distribution function of the standard Gaussian distribution. Lemma~\ref{lemmoments} implies that for all $\varepsilon \in ]0,1[$, there exists $T_{\varepsilon}$ such that for all $t \geq T_{\varepsilon}$,
\begin{equation}
\label{encadrement}
\frac{q!}{t^q}(1-\varepsilon) \leq \E \left( \left ( K_{\rho} \left ( 1+\frac{F_1}{t} \right )\right )^q\right ) \leq \frac{q!}{t^q}(1+\varepsilon).
\end{equation}
Furthermore, for $x \in \R$,
\begin{eqnarray*}
{\rm{P}}(S_n(E_{n-k+1,n}) \leq x ) - \Phi(x) & = & \int_0^{T_{\varepsilon}} ({\rm{P}}(S_n(t) \leq x ) - \Phi(x))h_n(t) dt \\
 \ & + &\int_{T_{\varepsilon}}^{\infty} ({\rm{P}}(S_n(t) \leq x ) - \Phi(x))h_n(t) dt =: A_n + B_n,
\end{eqnarray*}
where $h_n$ is the density of the random variable $E_{n-k+1,n}$. First, let us focus on the term $A_n$. We have,
\[ |A_n| \leq 2 {\rm{P}}(E_{n-k+1,n} \leq T_{\varepsilon}). \]
Since $E_{n-k+1,n}/\log(n/k) \toP 1$ (see~\cite{thetaWT}, Lemma~1),
 it is easy to show that $A_n \to 0$. Now, let us consider the term $B_n$. For the sake of simplicity, let us denote:
\[ \left \{ Y_i = K_{\rho} \left ( 1+\frac{F_i}{t} \right )- \mu_{\rho}(t), \ i=1,\ldots,k-1\right \}. \]
Clearly, $Y_1,\ldots,Y_{k-1}$ are independent, identically distributed and centered random variables. Furthermore, for $t \geq T_{\varepsilon}$,
\begin{eqnarray*}
\E (|Y_1|^3) & \leq & \E \left ( \left ( K_{\rho} \left ( 1+\frac{F_1}{t}\right )+\mu_{\rho}(t) \right )^3 \right ) \\
 \ & = & \E \left ( \left ( K_{\rho} \left ( 1+\frac{F_1}{t}\right ) \right )^3 \right ) +(\mu_{\rho}(t))^3 + 3\E \left ( \left ( K_{\rho} \left ( 1+\frac{F_1}{t}\right ) \right )^2 \right ) \mu_{\rho}(t) \\
 \ & + & 3\E \left ( K_{\rho} \left ( 1+\frac{F_1}{t}\right ) \right ) (\mu_{\rho}(t))^2 \\
  \ & \leq & \frac{1}{t^3}C_1(q,\varepsilon) < \infty,
\end{eqnarray*}
from~(\ref{encadrement}) where $C_1(q,\varepsilon)$ is a constant independent 
of $t$. Thus, from Esseen's inequality (see~\cite{pet75}, Theorem 3), we have:
\[ \sup_x| {\rm{P}}(S_n(t) \leq x)-\Phi(x)| \leq C_2L_n, \]
where $C_2$ is a positive constant and
\[ L_n = \frac{(k-1)^{-1/2}}{(\sigma_{\rho}(t))^3} \E(|Y_1|^3). \]
From (\ref{encadrement}), since $t \geq T_{\varepsilon}$,
\[ (\sigma_{\rho}(t))^2 = \E \left ( \left ( K_{\rho} \left ( 1+\frac{F_1}{t}\right ) \right )^2 \right ) - \left ( \E \left ( K_{\rho} \left ( 1+\frac{F_1}{t} \right ) \right ) \right )^2 \geq \frac{1}{t^2}C_3(\varepsilon), \]
where $C_3(\varepsilon)$ is a constant independent of $t$. Thus, 
$L_n \leq (k-1)^{-1/2}C_4(q,\varepsilon)$ where $C_4(q,\varepsilon)$ is a constant independent of $t$, and therefore
\[ |B_n| \leq C_4(q,\varepsilon) (k-1)^{-1/2} {\rm{P}}(E_{n-k+1,n} \geq T_{\varepsilon}) \leq C_4(q,\varepsilon) (k-1)^{-1/2} \to 0, \]
which concludes the proof.
\CQFD


\end{document}